\documentclass[conference]{IEEEtran}
\IEEEoverridecommandlockouts
\usepackage{cite}
\usepackage{graphicx}
\usepackage{subcaption}
\usepackage{textcomp}
\usepackage[hidelinks]{hyperref}
\usepackage{amsmath}
\usepackage{algorithm}
\usepackage{array}
\usepackage{multirow}
\usepackage{tabulary}
\usepackage{amsmath}
\usepackage{algpseudocode}
\usepackage{float}
\usepackage{stfloats}
\begin{document}

\title{Homomorphic Encryption-Enabled Federated Learning for Privacy-Preserving Intrusion Detection in Resource-Constrained IoV Networks}

\makeatletter
\newcommand{\linebreakand}{%
\end{@IEEEauthorhalign}
\hfill\mbox{}\par
\mbox{}\hfill\begin{@IEEEauthorhalign}
}
\makeatother
	
\author{\IEEEauthorblockN{Bui Duc Manh, Chi-Hieu Nguyen, Dinh Thai Hoang and Diep N. Nguyen}
     School of Electrical and Data Engineering, University of Technology Sydney, Australia. \\      
}

\maketitle

\begin{abstract}
This paper aims to propose a novel framework to address the data privacy issue for Federated Learning (FL)-based Intrusion Detection Systems (IDSs) in Internet-of-Vehicles (IoVs) with limited computational resources. In particular, in conventional FL systems, it is usually assumed that the computing nodes have sufficient computational resources to process the training tasks. However, in practical IoV systems, vehicles usually have limited computational resources to process intensive training tasks, compromising the effectiveness of deploying FL in IDSs. While offloading data from vehicles to the cloud can mitigate this issue, it introduces significant privacy concerns for vehicle users (VUs). To resolve this issue, we first propose a highly-effective framework using homomorphic encryption to secure data that requires offloading to a centralized server for processing. Furthermore, we develop an effective training algorithm tailored to handle the challenges of FL-based systems with encrypted data. This algorithm allows the centralized server to directly compute on quantum-secure encrypted ciphertexts without needing decryption. This approach not only safeguards data privacy during the offloading process from VUs to the centralized server but also enhances the efficiency of utilizing FL for IDSs in IoV systems. Our simulation results show that our proposed approach can achieve a performance that is as close to that of the solution without encryption, with a gap of less than 0.8\%. 
\end{abstract}


\section{Introduction}

In recent years, the Internet of Vehicles (IoV) has witnessed as a significant research field with the development of intelligent transport systems (ITSs), connected autonomous vehicles (CAV), vehicular ad hoc networks (VANET), and in-vehicle networks (IVNs)~\cite{ji2020iov}. Particularly, IoV is a combination of VANET and IoT, which is built based on the communication standards of VANET~\cite{lampre2023ids}. In the IoV network, the vehicles equipped with IoT sensors can transmit and exchange data via the roadside units (RSUs), thereby providing intelligent decision-making based on the collected data. However, the massive network connectivity in IoV leads to substantial concerns in cybersecurity~\cite{lampre2023ids}. Regarding cyber threats, the IoV networks are vulnerable to various attacks. For example, the attacker can impersonate one vehicle in the network stream to steal or inject false information into a vulnerable vehicle via a spoofing attack~\cite{amoo2015vul}. Additionally, traditional cyberattacks, such as denial of service (DoS), can severely impact IoV networks by overwhelming them with an intensive volume of traffic, thereby disrupting network services. Therefore, effective detection and defence solutions for cybersecurity in IoV networks are urgently needed to ensure the integrity, reliability, and safety of connected vehicle systems.

Machine Learning (ML) has emerged as a promising approach that can integrate with modern networks to form an intelligent intrusion detection system (IDS). Regarding IoV networks, the authors in~\cite{zer2021vtc} propose a Convolutional Long Short Term Memory Network (ConvLSTM) to detect anomalies in IoT sensors integrated CAVs. The simulation results show that the deep learning model can detect various anomalies in sensor data with an F1-score of 97\%. Moreover, in~\cite{all2021ids}, the authors evaluate various deep learning techniques to detect attacks in vehicular network traffic, which achieve accuracy from 92\% to nearly 99\%. However, traditional deep learning approaches operate based on centralized learning paradigms, which may not be efficient for the decentralized nature of IoV/IoT networks. Therefore, Federated Learning (FL) is a significant solution that allows deep learning models to learn with decentralized user-generated data~\cite{mcmahan2017fedavg}. Regarding IoT networks, the authors in~\cite{khoa2020fl} propose a collaborative framework that utilizes FL for intrusion detection in IoT networks. Experiment results using Deep Belief Network (DBN) and Deep Autoencoder (DAE) show the accuracy of detection from 93\% to 98\%. 

Despite the advantage of FL in IDSs, there are still some major challenges when deploying it in practical IoV networks. Specifically, in FL-based IDS in IoV networks, vehicles or RSUs often serve as workers to store and process all the learning tasks (e.g., training and classification)~\cite{ji2020iov}. However, in practice, both RSUs and vehicles usually have limited computing and storage resources, and thus, storing and processing learning tasks at RSUs and VUs are ineffective. It is important to note that in conventional FL processes, a delay from one computing node can cause a delay for the whole system~\cite{lim2020federated}. Therefore, several works propose solutions to upload data from RSUs and vehicles to powerful servers (e.g., centralized servers) for processing~\cite{yo2020hybrid} \cite{ji2021com}. This approach can be very effective in deploying ML algorithms as all the data is collected and processed at the centralized servers. However, it also raises a serious concern regarding the data privacy of VUs, as all the data is now stored and processed externally. 

To overcome the above challenges, we propose a novel privacy-preserving FL framework for intrusion detection in IoV networks. The proposed framework can effectively protect VUs' privacy and detect cyberattacks, given VUs' limited computational resources. Specifically, based on the current computing and storage resource capabilities, VUs can decide the amount of data they need to upload to the server for processing. To preserve the privacy of the VUs, this data will be encrypted by employing Homomorphic Encryption (HE) before being uploaded to the server. While this encryption method enhances data privacy, it presents significant challenges for the centralized server, which is tasked with training on the encrypted data offloaded from the VUs. To tackle this issue, we develop a robust training algorithm leveraging the Single Instruction Multiple Data (SIMD) and the bootstrapping capabilities of the underlying HE scheme. This enables direct computation on the quantum-secure encrypted ciphertexts without the need for decryption. This approach not only maintains the confidentiality of data during the offloading process from VUs to the centralized server but also boosts the efficiency of using FL for IDSs within IoV networks. Simulation results on real-world datasets show that our proposed framework achieves not only a high accuracy (approximately 91\%) in detecting attacks but also exhibits great performance, closely approaching the benchmark without using encryption (with a gap of less than 0.8\%).  

\begin{figure*}[t]
    \centering
    \begin{subfigure}[b]{0.5\textwidth}
        \centering
        \includegraphics[width=\linewidth]{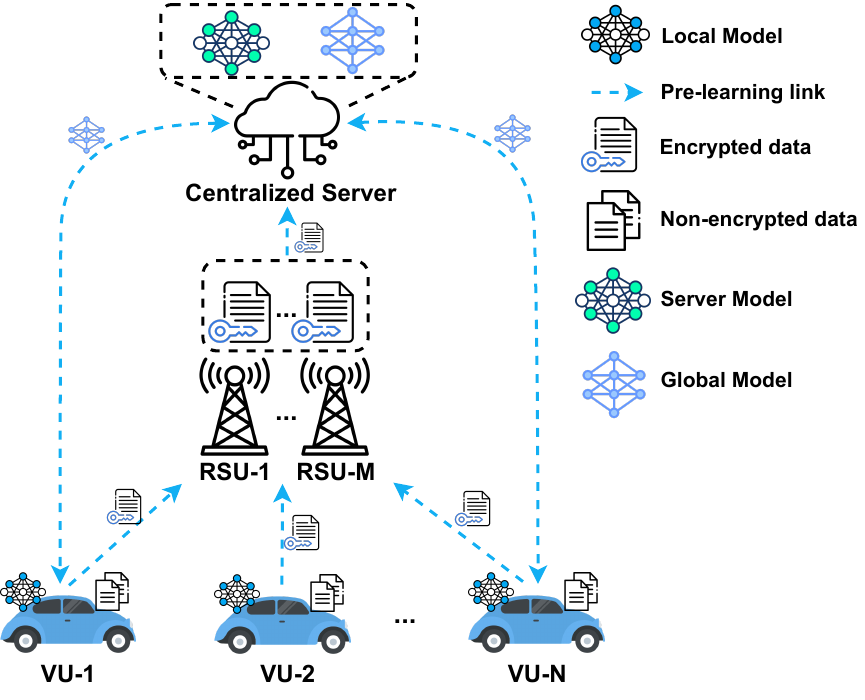}
        \captionsetup{justification=raggedright, singlelinecheck=false, margin=2cm}
        \caption{Pre-learning process}
        \label{fig:pre_model}
    \end{subfigure}%
    \hfill
    \begin{subfigure}[b]{0.5\textwidth}
        \centering
        \includegraphics[width=\linewidth]{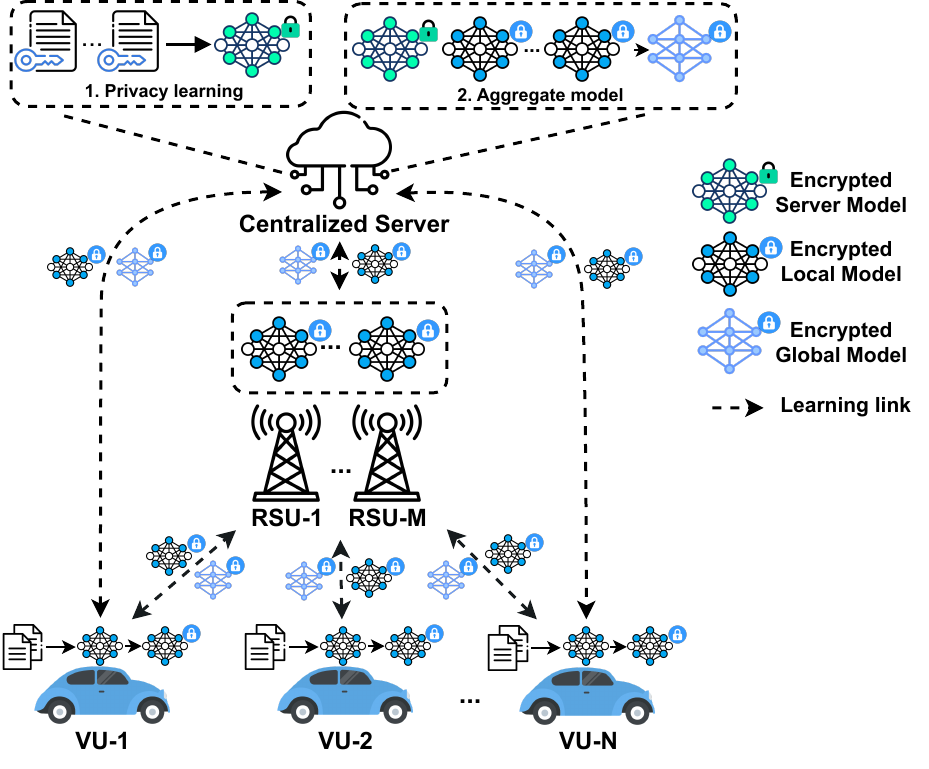}
        \captionsetup{justification=raggedright, singlelinecheck=false, margin=2cm}
        \caption{Privacy-preserving learning process}
        \label{fig:main_model}
    \end{subfigure}
    \caption{The proposed privacy-preserving intrusion detection framework including pre-learning and privacy-preserving learning.}
    \label{fig:sys}
\end{figure*}

\section{System Model}

The proposed system model is illustrated in Fig.~\ref{fig:sys}. The system consists of a centralized server (CS), $M$ RSUs and $N$ VUs. Initially, the VUs enter the pre-learning phase by assessing their computational resources and determining the optimal amount of data that can be processed locally. The rest of the data will be offloaded to the centralized server for processing. However, before offloading the data to the centralized servers, VUs will generate HE key pairs and use them to encrypt uploading data. The encrypted data will then be offloaded to the centralized server via RSUs, as illustrated in Fig.~\ref{fig:pre_model}. Upon receiving the offloaded encrypted data, the centralized server will compile it into an encrypted dataset. Using this dataset, two learning models will be developed: the server model and the global model, each serving distinct purposes. The server model will be utilized to train the encrypted dataset within the centralized server, whereas the global model will be distributed to the VUs for local training. Once the global model is sent to the VUs, the privacy-preserving learning process will commence.

The privacy-preserving learning process will be divided into different learning periods. During each learning period, each VU will use the global model to train on its local data. After completing the training, the VU will encrypt its trained model before sending it to the centralized server. Concurrently, the centralized server will train its encrypted data using our proposed CKKS scheme, detailed in Section~\ref{sec:3a}. Upon receiving all the encrypted trained models from the VUs, the centralized server will aggregate them to create a new global model (the aggregation method is detailed in Section~\ref{sec:3b}). This updated global model is then sent back to the VUs, and the next learning period begins. This process repeats until the global model converges or until a predefined number of learning periods has been completed.

\section{The Privacy-Preserving FL Framework for Encrypted Data}

\subsection{The Classification-based Deep Neural Network for Encrypted Data}
\label{sec:3a}

\begin{figure}[t]
    \centering
    \includegraphics[width=\linewidth]{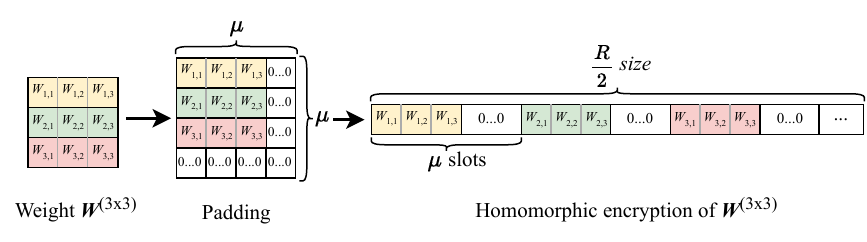}
    \caption{Encrypt process of a 3x3 matrix \textbf{\textit{W}}.}
    \label{fig:encW}
\end{figure}

To integrate HE with deep neural networks, we propose to use the Cheon-Kim-Kim-Song (CKKS) scheme. The reason is that it allows the encryption and calculation of real numbers, which is suitable for deep learning~\cite{cheon2017ckks}. The CKKS provides basic HE algorithms as follows~\cite{hieu2024enc}:

\begin{itemize}
    \item $SKGen(n)$: generate random secret key $s_n^{sk}$ for user $n$.
    \item $PKGen(s_n^{sk})$: create the public key $s_n^{pk}$ for user $n$ based on the secret key $s_n^{sk}$.
    \item $Enc(s_n^{pk}, z)$: encrypt vector $z$ into a ciphertext $\hat{z}$ by using the public key $s_n^{pk}$.
    \item $Dec(s_n^{sk}, \hat{z})$: decrypt vector $\hat{z}$ into its plain form $z$ by using the secret key $s_n^{sk}$
    \item $Add(\hat{z},\hat{x})$, $Sub(\hat{z},\hat{x})$, and $Mul(\hat{z},\hat{x})$: perform element-wise addition, subtraction and multiplication between two ciphertexts $\hat{z}$ and $\hat{x}$.
\end{itemize}


Specifically, HE schemes require a ring dimension $\textit{R}$, which maintains the security level, multiplication depth, and noise level~\cite{gentry2009fully}, thereby allowing accurate computations over encrypted data. Following that, to design a deep neural network for encrypted data, we employ the single instruction multiple data (SIMD) from the CKKS scheme, which packs multiple plaintexts into a single ciphertext. The size of ciphertext is denoted as $\textit{B}$, where $\textit{B}=\textit{R}/2$. Alternatively, the CKKS can encode and encrypt a square matrix of size at most $\mu \times \mu$, where $\mu = \lfloor\sqrt\textit{B}\rfloor$ by initially flattening it into a vector. This thus enables element-wise operation on the plaintext slots concurrently. For clarity, Fig.~\ref{fig:encW} describes the implementation of the weight matrix encryption method. Let $\phi_i$ denote the parameter of $i$ linear layer which $\phi_i = (W_i,b_i)$. The weight matrix $W^{u \times v}_i$ with $u$ and $v$ as the input and output dimensions of the layer is applied to the encoding process:
\begin{equation}
    Encode(W_i) = Flatten\big(Pad(W_i,0,\mu)\big) ,
\label{eq:encode}
\end{equation}

\noindent where the matrix $W^{u \times v}_i$ is first zero-padded to $W^{\mu \times \mu}_i$ with the size of ($\mu$, $\mu$) to fit within $\textit{B}$ size. The weight matrix is then flattened to form an encoded vector. After that, this encoded vector is padded to ensure its length equals half of the ring dimension~\cite{gentry2009fully}. After that, the encoded vector is encrypted by CKKS, which can be defined by:
\begin{equation}
    \hat{W_i} = Enc(s^{pk},Encode(W_i)) ,
\end{equation}

\noindent where $s^{pk}$ is the public key generated by the users. As a result, $\hat{W_i}$ is the encrypted weight of layer $i$, which can be used to operate with encrypted training data. Therefore, the output of the forward propagation over the $i$-th layer can be calculated as:
\begin{equation}
    \hat{x}_{i+1} = \hat{\sigma}\big(Add(Mul(\hat{x}_i,\hat{W}_i),\hat{b}_i)\big) ,
\end{equation}

\noindent where $\hat{W}_i$ and $\hat{b}_i$ are the encrypted weight and bias at layer $i$. Particularly, $\hat{\sigma}$ illustrates the polynomial approximation of the activation function $\sigma$ using the Chebyshev polynomial~\cite{lee2022cheb}. In the considered deep neural network, the \textit{Swish} (SiLU) activation function is chosen due to its advantage in solving the ``dying ReLU'' problems~\cite{elwing2018silu}. Subsequently, the encrypted output vectors consist of the distribution of the classes for classification tasks. It is noted that the \textit{Softmax} function is not applied in this work due to the exponential and inverse functions contained, which are non-homomorphic~\cite{hieu2024enc}.

In the backpropagation process, we apply the Stochastic Gradient Descent (SGD) for mini-batch regarding the optimization. After calculating the encrypted gradients for each layer, the SGD update of the encrypted weight can be formulated as:
\begin{equation}
    \hat{W_{i}} \leftarrow BootStrap\Big(Sub\Big(\hat{W_{i}},Mul\big(\eta,\frac{\partial \hat{L}}{\partial \hat{W_i}}\big)\Big)\Big)
\end{equation} 

\noindent where $\frac{\partial \hat{L}}{\partial \hat{W_i}}$ is the calculated encrypted gradient of $\hat{W_i}$, which is computed via the derivative of encrypted loss function $\hat{L}$. After the SGD update, the encrypted weight $\hat{W_i}$ is applied to the BootStrap method, which renews the ciphertext, allowing additional computation on $\hat{W_i}$ and reduces the magnitude of accumulated noise~\cite{gentry2009fully}.

\subsection{The Proposed FL Implementation}
\label{sec:3b}

In the pre-learning phase, each VU-$n$ evaluates its computing resources and chooses $p_n$\% of data to offload. After that, they generate a key pair, including a secret key $s^{sk}_n$ and a public key $s^{pk}_n$. In particular, the VU-$n$ divide its collected dataset $\mathcal{D}_n$ into the local dataset $\mathcal{DR}_n$ and offloaded dataset $\mathcal{DS}_n$ based on effective computing resources of the vehicles. The $\mathcal{DS}_n$ is then encrypted to $\hat{\mathcal{DS}}_n$ to protect the user data. The encrypted data is sent to the CS and combined to form an encrypted dataset $\hat{\mathcal{DS}}$. After that, the CS initializes the non-encrypt global model $\mathcal{M}_g$ and non-encrypt server model $\mathcal{M}_s$, then distributes $\mathcal{M}_g$ to each VU for local training. Subsequently, the public key is used to initialize the encrypted learning model $\hat{\mathcal{M}}_s$ and encrypted global model $\hat{\mathcal{M}}_g$ on the server.

Regarding the privacy-preserving learning phase, we consider $T$ learning rounds. At each round $\tau$, privacy is maintained by the non-encrypted training from the local learning model of VU-$n$ and encrypted training from the privacy-preserving learning model. After finishing the $\tau^\text{th}$ local training round, VU-$n$ encrypts the trained parameters $\hat{\phi}_{\tau}^n$ and sends them to the CS. The CS retrieves the encrypted parameters from $\hat{\mathcal{M}}_s$ along with the local encrypted parameters and aggregates by the FedAvg algorithm for encrypted data, which can be defined by:

\begin{equation}
\hat{\phi}_g^{(\tau+1)} = Mul\Big(\frac{1}{N+1}, Add\Big(\sum_{n=1}^{N}\hat{\phi}_{n}^\tau,\hat{\phi}_{s}^\tau\Big)\Big) .
\end{equation}

The global parameters $\hat{\phi}_g^{(\tau+1)}$ are then updated to the encrypted global model $\hat{\mathcal{M}}_g$ and sent back to the VUs, which is then decrypted by the VUs for the next learning round. The learning process continues until the global model converges and obtains the optimized parameters. 

\begin{algorithm}
\caption{Proposed Privacy-Preserving FL Framework}
\label{ago: ago1}
\begin{algorithmic}[1]
\For{$\forall n \in N$}
    \State Calculate $p_n$\% for offloading
    \State \parbox[t]{0.8\linewidth}{Generate a secret key and a public key: $s_n^{sk} = SKGen(n)$ and $s_n^{pk} = PKGen(s_n^{sk})$}
    \State \parbox[t]{0.8\linewidth}{Split the dataset $\mathcal{D}_n$ into  $\mathcal{DR}_n$ and  $\mathcal{DS}_n$ where  $\mathcal{DS}_n = \mathcal{D}_n \times p_n $ and $\mathcal{DR}_n = \mathcal{D}_n - \mathcal{DS}_n$}
    \State Generate the encrypted data $\hat{\mathcal{DS}_n} = Enc(s_n^{pk},\mathcal{DS}_n)$
    \State Send encrypted data $\hat{\mathcal{DS}_n}$ to the CS
\EndFor
\State CS combines the received encrypted data $\hat{\mathcal{DS}_n}$ into $\hat{\mathcal{DS}}$
\State CS initializes the $\mathcal{M}_g$ and $\mathcal{M}_s=\mathcal{M}_g$
\State Transmit $\mathcal{M}_n$ to $N$ VUs which $\mathcal{M}_n = \mathcal{M}_g$
\State Generate encrypted model $\hat{\mathcal{M}_s}$ and $\hat{\mathcal{M}_g}$ via $s_n^{pk}$ where $\phi_n = Dec(\hat{\phi_n})$
\While{$\tau \leq T_{\max}$ or training process does not converge}
    \State $\hat{\mathcal{M}_s}$ learns the encrypted data $\hat{\mathcal{DS}}$
    \State $\hat{\mathcal{M}_s}$ produces encrypted parameters $\hat{\phi}_{s}^\tau$
    \For{$n \in N$}
        \State $\mathcal{M}_n$ learns the local data  $\mathcal{DR}_n$
        \State Calculate local parameters $\phi_{\tau}^n$
        \State Encrypt local parameters $\hat{\phi}_{\tau}^n = Enc(s_n^{pk},\phi_{\tau}^n)$
        \State Send local encrypted parameters to the CS.
    \EndFor
    \State \parbox[t]{0.8\linewidth}{The CS calculates and produces the encrypted global model $\hat{\phi}_g^{(\tau+1)}$.}
    \State Send the updated global model $\hat{\phi}_g^{(\tau+1)}$ back to $N$ VUs
    \For{$\forall n \in N$}
        \State Decrypt the model $\phi_g^{(\tau+1)} = Dec(s_n^{sk},\hat{\phi}_g^{(\tau+1)})$
    \EndFor
\EndWhile
\State Predict $\hat{Y}_n$ based on the encrypted training data $\hat{X}_n$ at each VU-n and optimal global model $\phi^*$.
\end{algorithmic}
\end{algorithm}

In summary, the learning process of the privacy-preserving intrusion detection framework for IoV is described in Algorithm~\ref{ago: ago1}.

\section{Performance Evaluation}

\subsection{Simulation Setup}

In this section, the proposed privacy-preserving model is validated on the real-world dataset of network traffic attacks on IoT devices, named Edge-IIoT dataset~\cite{2022edgeiiot}. The Edge-IIoT dataset includes 20 million raw normal traffic and attack traffic collected from 13 IoT devices. Attacks can be grouped into the five most common types, including Distributed Denial of Service (DDoS), Injection, Man-in-the-Middle (MitM), Malware, and Reconnaissance. After applying downsample and oversample to overcome the imbalance, the dataset includes 31,400 samples with details presented in Table~\ref{tab:num_sample}. Subsequently, the dataset is divided into training and testing sets (80\%-20\%). The training and testing sets are then nominalized and scaled within the range of (0,1). Regarding the neural network, we design a fully connected network consisting of an input layer, 2 hidden layers, and an output layer. The respective layers contain 32, 16, 16, and 6 neurons. Apart from the input layer, each layer is attached to the SiLU activation function. 

\begin{table}[t]
\centering
\caption{The distribution of classes of the dataset}
\label{tab:num_sample}
\begin{tabular}{|c|c|}
\hline
\textbf{Class} & \textbf{Number of samples} \\ \hline
Normal & 5,320 \\ \hline
DDoS & 5,472 \\ \hline
MitM  & 4,000 \\ \hline
Injection & 5,589 \\ \hline
Malware & 5,504 \\ \hline
Reconnaissance & 5,515 \\ \hline
Total & 31,400 \\ \hline
\end{tabular}
\end{table}

During the simulation, we assume that the collected dataset of each VU has the same class distribution. Similar to~\cite{ji2021com}, we consider the approach to offload partial data to the server as the benchmark for our proposed framework. Nevertheless, it is noted that in such a benchmark, the FL framework does not consider the privacy of the VUs. In this scenario, we train the non-encrypted data using the non-encrypted model at both CS and VUs. Consequently, the trained global model is employed to evaluate the accuracy of our proposed framework and other benchmarks. In our experiment setup, the proposed framework consists of 2 VUs and 3 VUs, which can send 10\% and 20\% of their local data. 


\subsection{Evaluation Metrics}

To evaluate the performance of the detection model, the confusion matrix is utilized, which is suitable for a machine learning-based classification system~\cite{fawcett2006roc}. We denote TP, TN, FP, and FN as ``True Positive'', ``True Negative'', ``False Positive'', and ``False Negative''. Assuming the system consists of $C$ classes, which include normal and attack traffic, the accuracy can be calculated as:
\begin{equation} 
    \text{$Accuracy$} = \frac{1}{C} \sum_{c=1}^C \frac{TP_c + TN_c}{TP_c + TN_c + FP_c + FN_c}.
\end{equation}

The macro-average precision and recall are utilized in this term. Given $K$ as the number of classes in the system, the macro-average precision is:
\begin{equation}
    \text{$Precision$} = \frac{1}{K} \sum_{k=1}^K \frac{TP_k}{TP_k + FP_k}.
\end{equation}

The macro-average recall is calculated as follows:
\begin{equation}
    \text{$Recall$} = \frac{1}{K} \sum_{k=1}^K \frac{TP_k}{TP_k + FN_k}.
\end{equation}

\begin{figure}[t]
    \centering
    \begin{subfigure}[b]{0.24\textwidth}
        \centering
        \includegraphics[width=\linewidth]{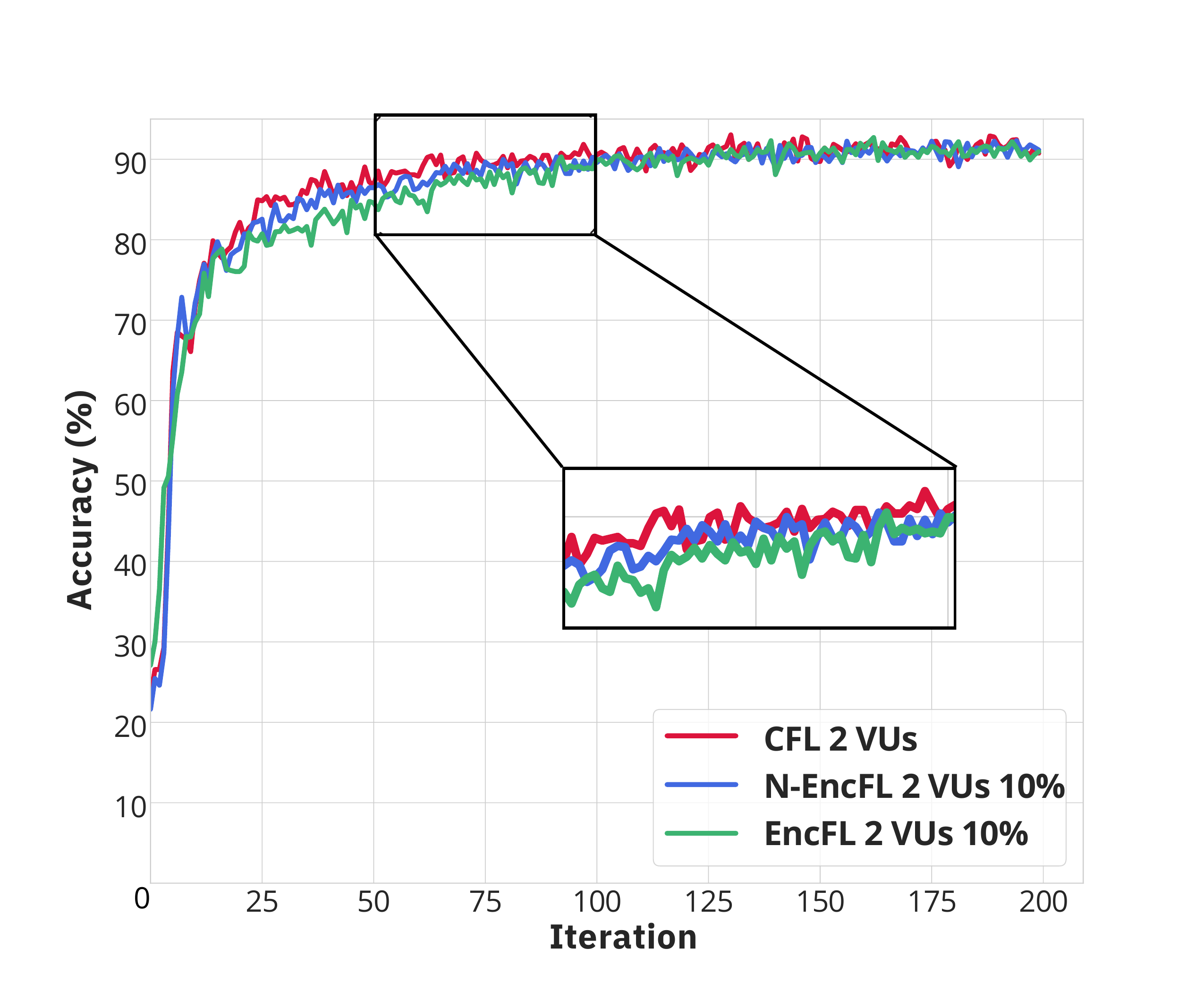}
        \caption{2 VUs with 10\% offload data}
        \label{fig:enc_convergence2w}
    \end{subfigure}
    \hfill
    \begin{subfigure}[b]{0.24\textwidth}
        \centering
        \includegraphics[width=\linewidth]{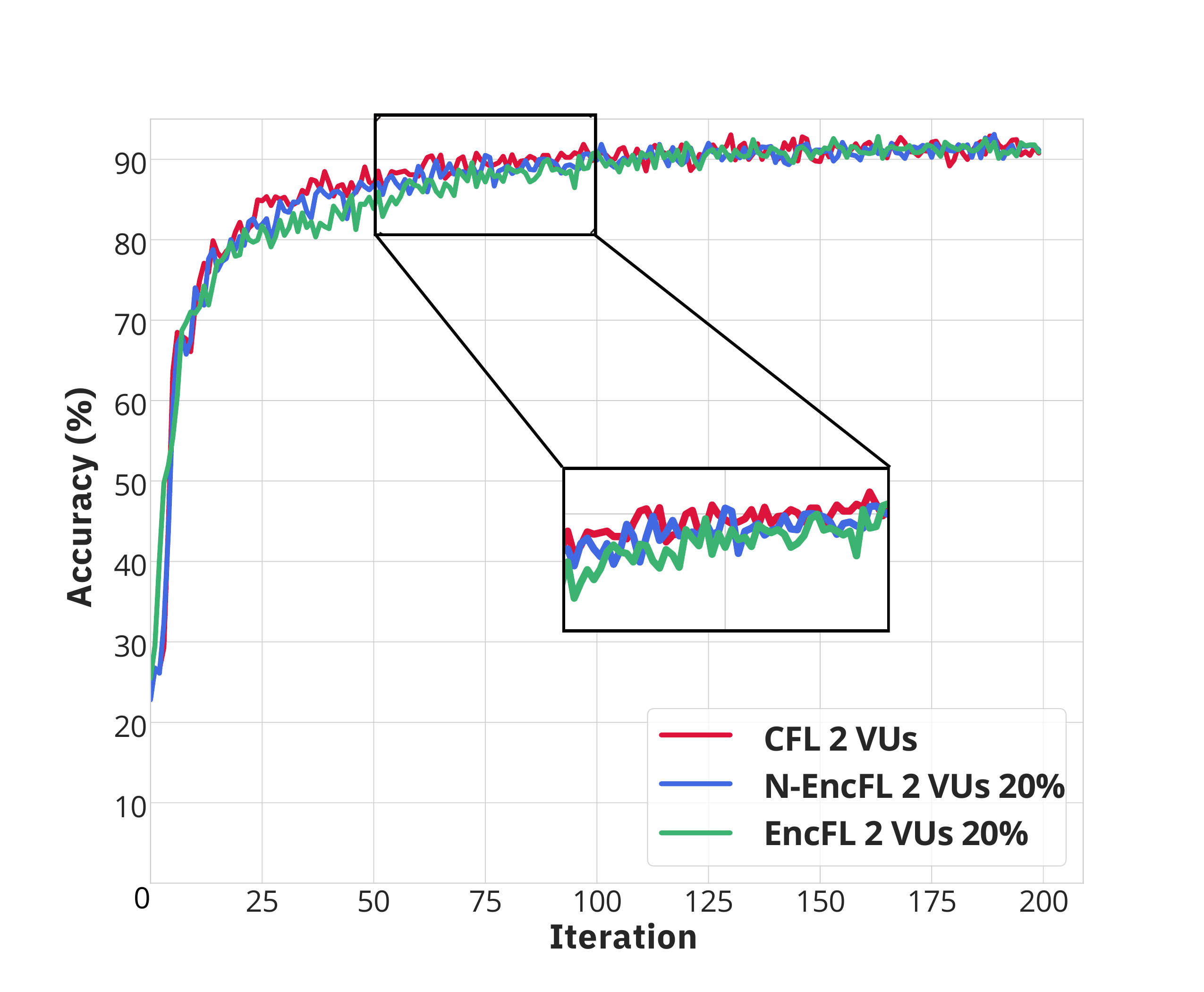}  
        \caption{2 VUs with 20\% offload data} 
        \label{fig:enc_convergence2w_20}
    \end{subfigure}
    \hfill
    \begin{subfigure}[b]{0.24\textwidth}
        \centering
        \includegraphics[width=\linewidth]{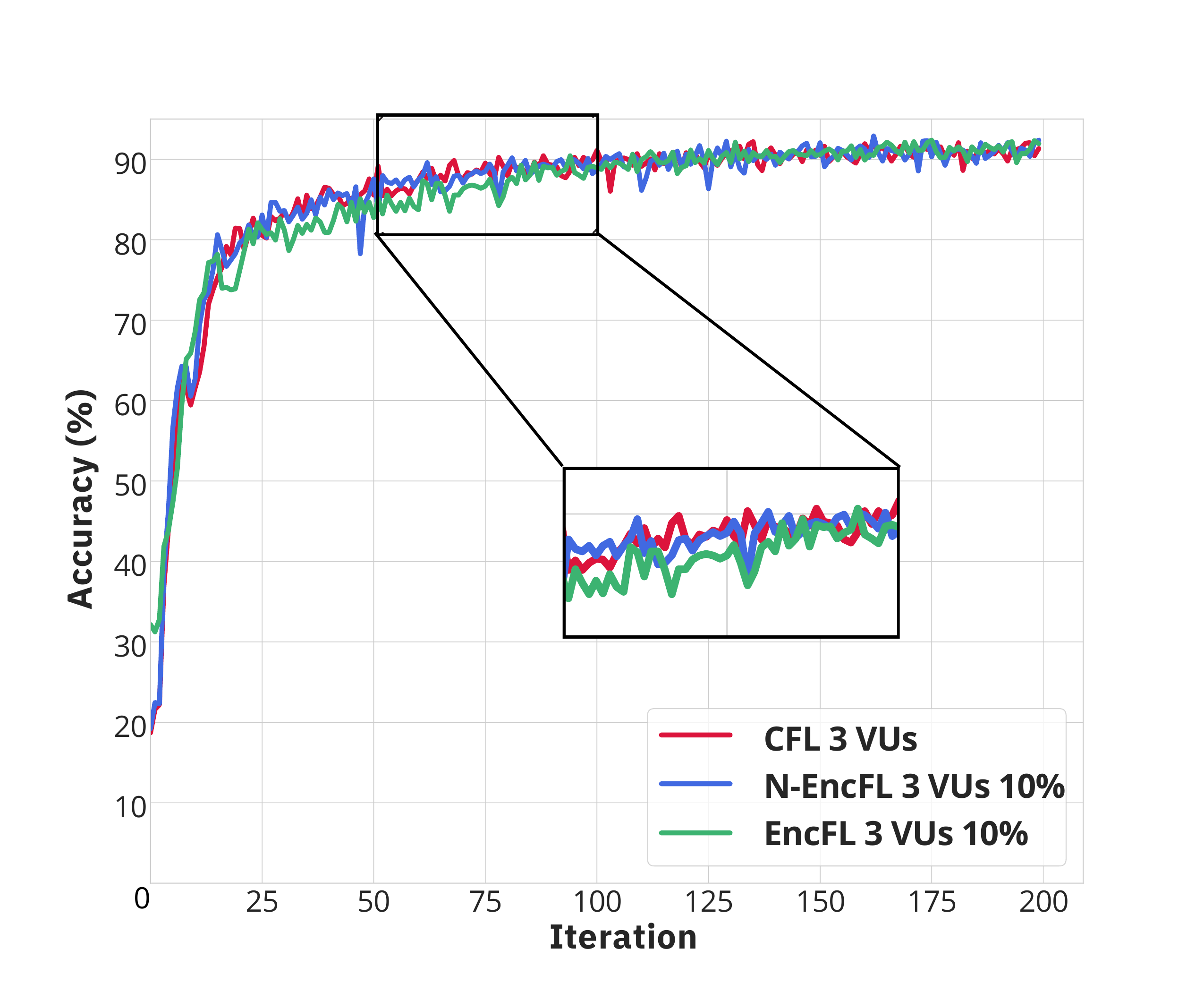}
        \caption{3 VUs with 10\% offload data}
        \label{fig:enc_convergence3w_10}
    \end{subfigure}
    \hfill
    \begin{subfigure}[b]{0.24\textwidth}
        \centering
        \includegraphics[width=\linewidth]{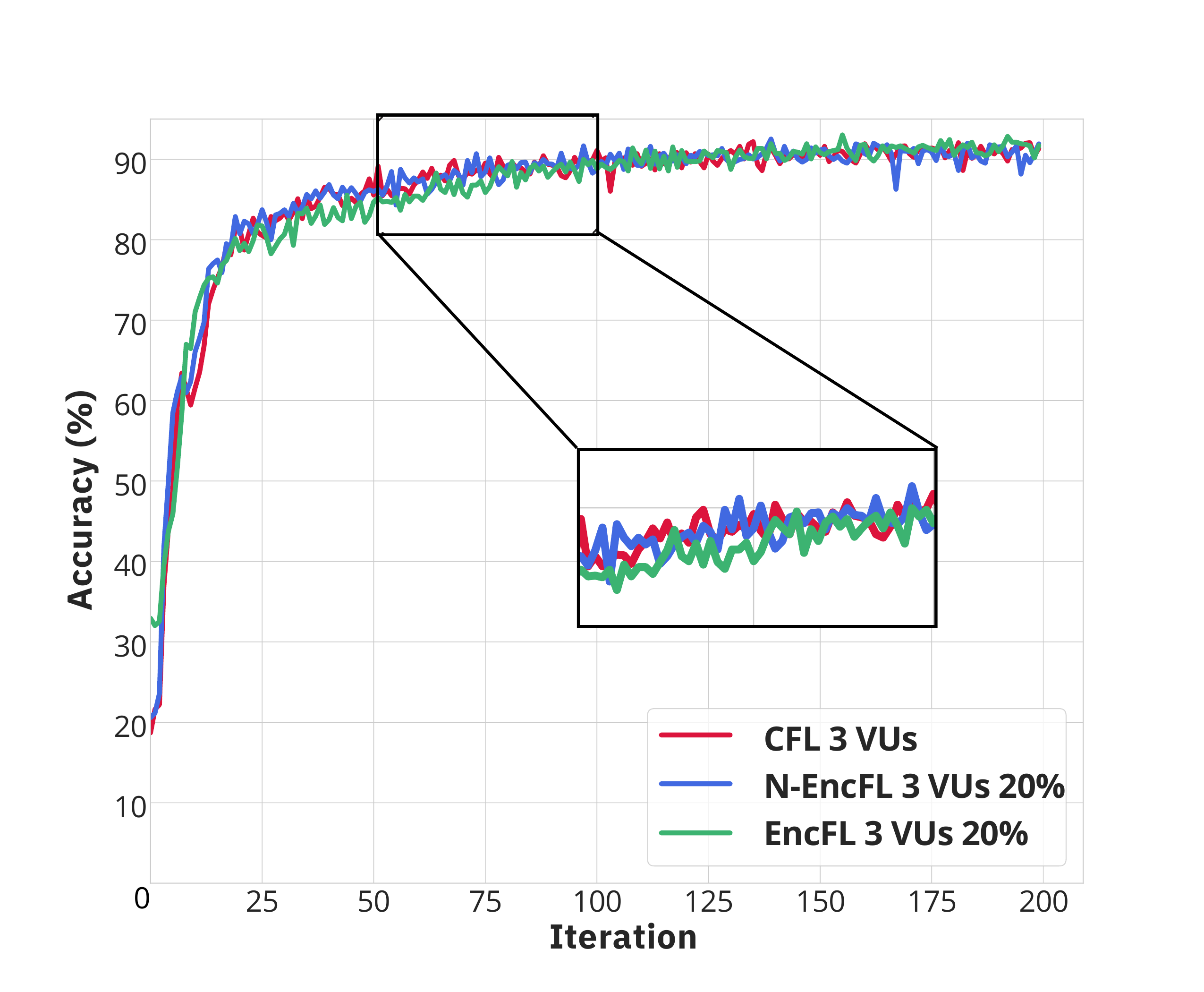}
        \caption{3 VUs with 20\% offload data}
        \label{fig:enc_convergence3w_20}
    \end{subfigure}
    \caption{Convergence of privacy-preserving learning with VUs.}
    \label{fig:enc_convergence}
\end{figure}

\begin{table*}[htbp]
\centering
\caption{Simulation results}
\label{tab:performance_metrics}
\begin{tabulary}{\textwidth}{|C|C|C|C|C|C|C|C|C|C|C|}
\hline
\multirow{2}{*}{\textbf{Model}} & \multicolumn{4}{c|}{\textbf{2 Vehicle Users}} & \multicolumn{4}{c|}{\textbf{3 Vehicle Users}} \\ \cline{2-9} 
                                & \multicolumn{2}{c|}{\textbf{N-EncFL}} & \multicolumn{2}{c|}{\textbf{EncFL}} & \multicolumn{2}{c|}{\textbf{N-EncFL}} & \multicolumn{2}{c|}{\textbf{EncFL}} \\ \cline{2-9}
                                & \textbf{10\% data} & \textbf{20\% data} & \textbf{10\% data} & \textbf{20\% data} & \textbf{10\% data} & \textbf{20\% data} & \textbf{10\% data} & \textbf{20\% data} \\ \hline
Accuracy       & 91.728         & 91.806         & 91.173         & 91.142         & 91.806         & 91.744         & 90.926         & 91.049         \\ \hline
Precision      & 92.767         & 92.868         & 92.319         & 92.254         & 92.787         & 92.730         & 91.992         & 92.161         \\ \hline
Recall         & 91.875         & 91.930         & 91.360         & 91.322         & 91.931         & 91.870         & 91.118         & 91.234         \\ \hline
\end{tabulary}
\end{table*}


\begin{figure}[t]
    \centering
    \begin{subfigure}[b]{0.48\textwidth}
        \centering
        \includegraphics[width=0.84\linewidth]{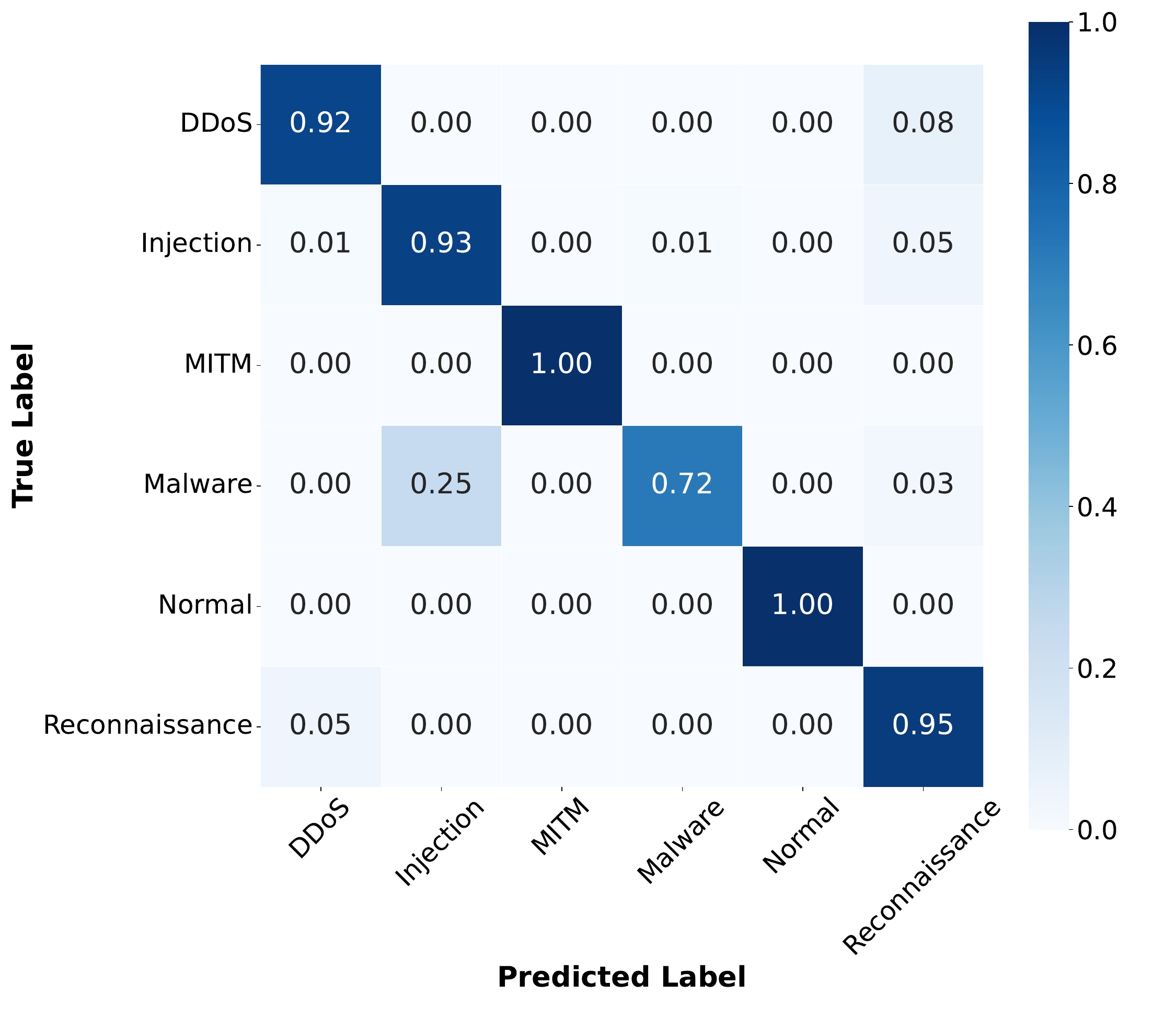}
        \caption{Offloading with 10\% non-encrypt data.}
        \label{fig:flraw10_2w_cm}
    \end{subfigure}%
    \hfill
    \begin{subfigure}[b]{0.48\textwidth}
        \centering
        \includegraphics[width=0.84\linewidth]{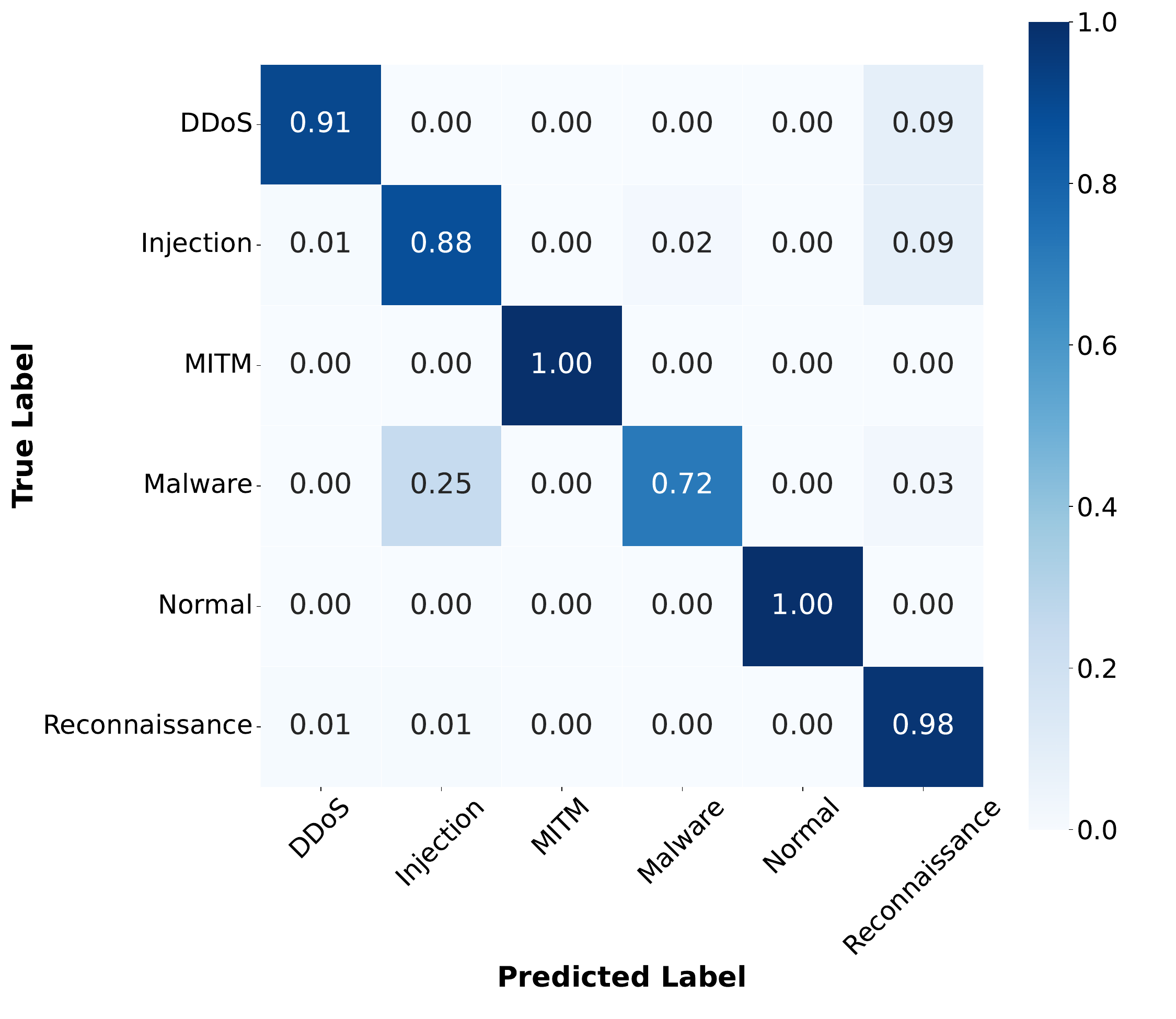}
        \caption{Offloading with 10\% encrypted data.}
        \label{fig:flraw20_2w_cm}
    \end{subfigure}
    \caption{Classification results of 2 VUs}
    \label{fig:confusion_matrices2}
\end{figure}

\subsection{Simulation Results}

\begin{figure}[t]
    \centering
    \begin{subfigure}[b]{0.48\textwidth}
        \centering
        \includegraphics[width=0.84\linewidth]{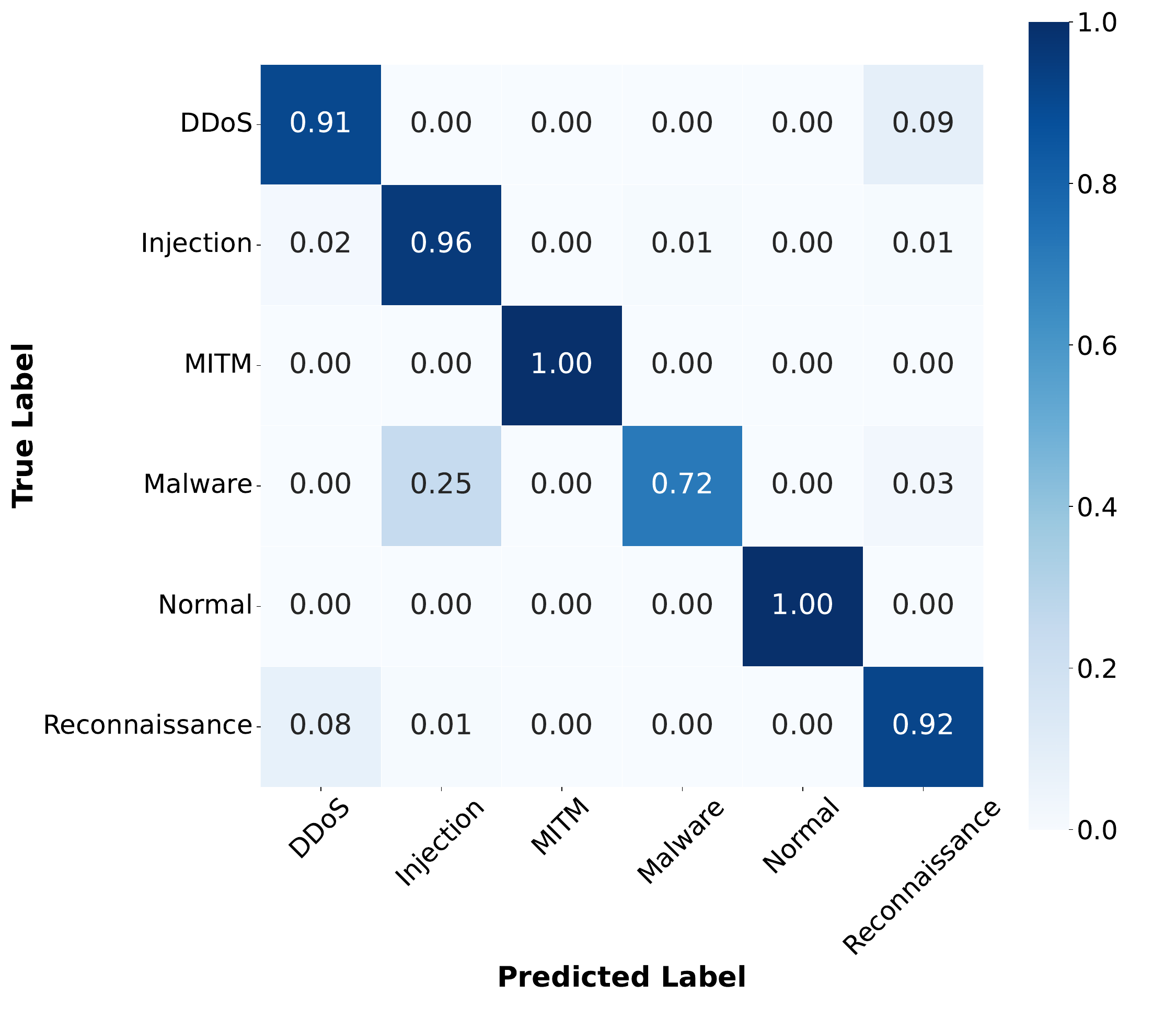}
        \caption{Offloading with 20\% non-encrypt data.}
        \label{fig:flraw20_3w_cm}
    \end{subfigure}%
    \hfill
    \begin{subfigure}[b]{0.48\textwidth}
        \centering
        \includegraphics[width=0.84\linewidth]{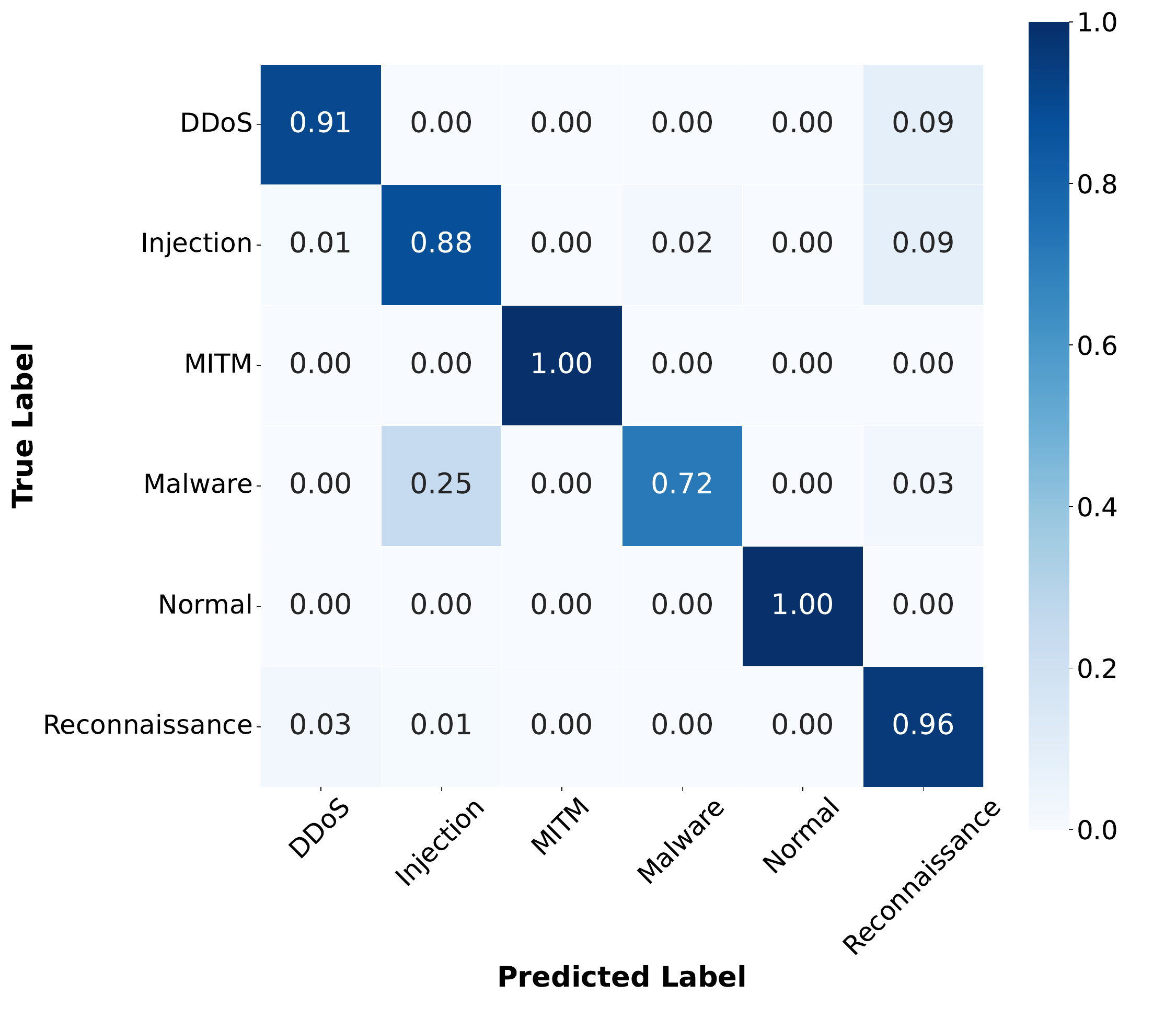}
        \caption{Offloading with 20\% encrypted data.}
        \label{fig:flenc20_3w_cm}
    \end{subfigure}
    \caption{Classification results of 3 VUs}
    \label{fig:confusion_matrices3}
\end{figure}

\subsubsection{Convergence Analysis} Fig.~\ref{fig:enc_convergence} illustrates the convergence of learning processes from three approaches, including conventional FL (CFL), FL with non-encrypted offloaded data (N-EncFL) and the proposed privacy-preserving learning (EncFL). As observed in Fig.~\ref{fig:enc_convergence2w} and Fig.~\ref{fig:enc_convergence2w_20} with 2 VUs, the CFL converges after 70 iterations, while the N-EncFL and EncFL require nearly 100 iterations to reach the convergence. Specifically, due to the different amounts of data handled by the VU, the CFL demonstrates a slightly better convergence rate compared to other approaches. Despite the trivial difference in convergence, the accuracy during the learning process of the three approaches remains nearly identical, stabilizing at approximately 92\%. Additionally, Fig.~\ref{fig:enc_convergence3w_10} and Fig.~\ref{fig:enc_convergence3w_20} describe the convergences in the scenarios with 3 VUs. Although the N-EncFL and traditional methods converge at nearly the same time, the EncFL require over 100 iterations to reach the convergence, which is slightly longer than other approaches. However, the gap in learning rate, which is about 15 to 20 iterations, is trivial. It is worth noting that the accuracy of EncFL remains consistent with that of the N-EncFL and  CFL, regardless of whether the amount of offloaded data is different. As a result, the proposed framework, which operates on encrypted data, achieves the same accuracy as those of the other benchmarks, i.e., N-EncFL and CFL.

\subsubsection{Performance Evaluation} Table~\ref{tab:performance_metrics} describes the performance in detecting attacks of two and three VUs in the IoV network. Overall, the accuracy, precision and recall of the two scenarios remain nearly identical. Regarding the different amounts of offloaded data, the results for N-EncFL and EncFL are close to those of other methods. Specifically, even when the data sent is 10\% or 20\%, the N-EncFL with two or three VUs achieves an accuracy of approximately 91.8\%. A similar trend is observed with EncFL, where the accuracy remains consistent regardless of the varying amounts of data offloaded. When comparing the results of EncFL and N-EncFL, we can observe that the accuracy, precision and recall of EncFL are slightly lower than those of the N-EncFL. In detail, the gap between N-EncFL and EncFL with two VUs is from 0.5\% to 0.6\%. For instance, with 10\% offloaded data, EncFL achieves an accuracy of 91.173\%, which is 0.55\% less than N-EncFL's 91.728\%. Additionally, the results of the three VUs show a similar pattern, with EncFL performing 0.6\% to 0.8\% less than RawFL. However, as observed in Fig.~\ref{fig:confusion_matrices2} and Fig.~\ref{fig:confusion_matrices3}, the overall accuracy of 6 classes is nearly the same. The differences primarily lie in the detection of ``Injection'' and ``Reconnaissance'', which accounts for the small gap between N-EncFL and EncFL. Although the accuracy of the ``Injection'' class of EncFL is less than N-EncFL, EncFL still achieves an 88\% detection rate accuracy for the ``Injection" attack. As a result, the small gap between N-EncFL and EncFL is acceptable, demonstrating that EncFL can classify each class with a high detection rate.

    

\section{Conclusion}
In this paper, we have proposed a novel privacy-preserving FL framework for intrusion detection in IoVs with limited computing resources. The proposed framework enables users to offload data to a centralized server, addressing the computational challenges during local training of the vehicles. To ensure user privacy, homomorphic encryption (HE) is applied to the data before offloading it to the server. The encrypted data is then processed by the training algorithm-based HE, which allows the server to learn from the encrypted data without knowing its content. The proposed framework can protect the privacy of users during the learning process, facilitating the efficient deployment of FL for IDSs in practical IoV networks. The simulation results show that our proposed framework can accurately detect cyberattacks in IoV networks. Although the accuracy of the encrypted neural network is slightly less than that of the raw ones, the gap is acceptable and can be optimized in future works.

\bibliographystyle{IEEEtran}
\bibliography{ref}

\end{document}